\documentclass[12pt,preprint]{aastex}
\received{}
\revised{}
\accepted{}

\shorttitle{\ion{O}{8} Absorption}
\shortauthors{Fang et al.}

\begin{document}

\title{{\sl Chandra} Detection of \ion{O}{8} Ly$\alpha$ Absorption from an Overdense Region in the Intergalactic Medium}

\author{Taotao Fang, Herman L.~Marshall, Julia C.~Lee, David S.~Davis AND\\Claude R.~Canizares}
\affil{Department of Physics and Center for Space Research, MIT\\NE80-6081, 77 Mass. Ave., Cambridge, MA 02139, fangt@space.mit.edu}

\begin{abstract}
We report the first detection of an \ion{O}{8} Ly$\alpha$ absorption
line associated with an overdense region in the intergalactic medium
(IGM) along the sightline towards PKS~2155-304 with the {\sl Chandra}
Low Energy Transmission Grating Spectrometer (LETGS). The absorption line is detected at $4.5\sigma$ level with $cz
\approx 16,600~\rm km\ s^{-1}$. At the same velocity \citet{sps98} detected a small group of spiral galaxies (with an
overdensity of $\delta_{gal} \sim 100$) and low metallicity \ion{H}{1}
Ly$\alpha$ clouds.  We constrain the intragroup gas that gives rise
to the \ion{O}{8} Ly$\alpha$  line to a baryon density in the range
$1.0 \times 10^{-5} < n_{b} < 7.5\times 10^{-5}~\rm cm^{-3}$ ($50 <
\delta_{b} < 350$) and a temperature of $4-5\times10^{6}$ K, assuming 0.1 solar abundance. These estimates are in accordance
with those of the warm/hot intergalactic medium (WHIM) that are
predicted from hydrodynamic simulations.  Extrapolating from this
single detection implies a large fraction of the ``missing baryons''
($\sim$ 10\%, or $\sim$ 30-40\% of the WHIM) are probed by the \ion{O}{8} absorber.

\end{abstract}


\section{Introduction}

The cosmic baryon budget at low and high redshift indicates that a large fraction of baryons in the local universe have so far escaped detection (e.g., \citealp{fhp98}). While there is clear evidence that a significant fraction of these ``missing baryons'' (between 20-40\% of total baryons) lie in photoionized, low-redshift Ly$\alpha$ clouds \citep{pss00}, the remainder could be located in intergalactic space with temperatures of $10^{5}-10^{7}$ K (warm-hot intergalactic medium, or WHIM). Resonant absorption from highly-ionized ions located in the WHIM gas has been predicted based on both analytic
studies of structure formation and evolution
(\citealp{sba81,aem94,plo98,fca00}) and cosmic hydrodynamic
simulations (\citealp{hgm98,cos99a,dco01,fbc02}). Recent discovery of \ion{O}{6}
absorption lines by the Hubble Space Telescope ({\sl HST}) and the Far
Ultraviolet Spectroscopic Explorer ({\sl FUSE}) (\citealp{tsa00,
tsj00, tgs01}) indicates that there may be a significant reservoir of
baryons in \ion{O}{6} absorbers. While Li-like \ion{O}{6} probes about
$\sim 30-40\%$ of the WHIM gas (\citealp{cto01, fbr01}),
the remaining $\sim 60-70\%$ is hotter and can only be probed by
ions with higher ionization potentials, such as H- and He-like Oxygen.

The WHIM gas, predicted by hydrodynamic simulations, typically has an
overdensity of $\delta_{b} = 5-200$ and is distributed in small groups
of galaxies or in large scale filamentary structures that connect
virialized halos. Given the expected physical conditions, current
instruments such as {\sl Chandra} and {\sl XMM} should be capable of
detecting resonance features from the WHIM gas \citep{fbc02}. A first
attempt by \citet{fmb01} with the {\sl Chandra} High Energy
Transmission Grating Spectrometer (HETGS) yielded an upper limit of ${\rm N(O~VIII) < 10^{17}\
cm^{-2}}$. Recently, with the {\sl Chandra} LETGS-HRC, \citet{nic02}
discovered X-ray resonant absorption features from warm/hot local gas
along the line of sight towards PKS~2155-304, giving the first X-ray
evidence for possible WHIM gas.  These absorbers are near zero
redshift and therefore most likely  associated with the Galaxy or
local group (see \S~2).

We report the first detection of an \ion{O}{8}
Ly$\alpha$ absorption line along the sightline towards PKS~2155-304 at
$cz \sim 16,600~\rm km~s^{-1}$ plausibly associated with an {\it
intervening}, modestly overdense region in the IGM. The observations were made with the {\sl
Chandra} LETG-ACIS (Advanced CCD Imaging Spectrometer).  Physical diagnostics show the gas that gives rise
to this resonance line has the typical properties of the WHIM,
indicating the detection of a significant fraction of the ``missing
baryons''.

\section{Data Analysis}

PKS~2155-304 at $z \sim 0.116$ \citep{bbc84} is one of the
brightest soft X-ray sources (e.g.  \citealp{umu82} and
references therein) and has been used as a {\sl Chandra} calibration
target. It was observed with the {\sl Chandra} LETG-ACIS (see, e.g., {\sl Chandra} Proposers' Observatory Guide at http://asc.harvard.edu/) on May 31,
2000 , December 6, 2000 and November 31, 2001. We restrict all spectra to the range 2---42 $\rm\AA$ to avoid complexities at longer wavelengths from the instrumental C-K edge. The events were summed in 0.025$\rm\AA$ bins, half of the instrumental line response function, which has nearly constant FWHM $\sim 0.05\,$~\AA\, across its bandpass (for data processing procedures, see \citealp{mcc01}). We corrected the wavelength scale by calibrating the spectrum with two known features (atomic \ion{O}{1} from the Galaxy and solid-state \ion{O}{1} from the instruments). 
The three data sets were separately
divided by the best fit Galactic absorbed power-law to remove the
underlying broad-band continuum behavior. The remaining
residuals were then fit with a five-order polynomial. Such a technique removes spectral features larger than $\sim
8{\rm \AA}$ (e.g. 5-10\% calibration uncertainties) but preserves narrow line
features. The data
from the three observations were then combined as a weighted sum that
accounts for the differences in exposures and effective area.

The observation during November 2001 has the strongest soft flux (
respectively for May 2000, December 2000 and November 2001, $F_{\rm 0.1-2.4} = 1.35, 0.87~{\rm and}~2.92 \times 10^{-10}~{\rm ergs~cm^{-2}s^{-1}}$). All continua are well  described
by a single power law ($\Gamma \sim 2.31 \pm 0.25, 2.60 \pm 0.20$,\ and\ $2.72 \pm 0.47$, respectively) absorbed by a Galactic hydrogen column of $N_{H}
= 1.36\times 10^{20} {\rm cm^{-2}}$ \citep{lsa95} (errors are quoted at 90\% confidence).

After a blind search for any statistically significant absorption features, two absorption features with S/N~$>$~4 were detected in the
2--42~\AA\, region of the LETGS spectral bandpass (Figure~1).  These
features  were subsequently fit in ISIS (Interactive Spectral
Interpretation System : http://space.mit.edu/ASC/ISIS/)
with Gaussian models (Table~1). The absorption
feature at $\sim 21.6$~\AA\, was reported by
\citet{nic02} in the LETGS-HRC archival data, which they attribute to \ion{O}{7}~He$\alpha$ resonance absorption from warm/hot local gas. We also show in Table~1 our results for two other features claimed by \citet{nic02}. The feature attributed to \ion{O}{8}~Ly$\alpha$ is marginally detected ($3\sigma$) while \ion{O}{7}~He$\beta$ is not significant ($1\sigma$). We defer the discussion of these three features to a later paper, noting only that they are at $cz\sim 0$ and  therefore probably associated with the Galaxy or local group. We searched for the 20\AA~feature in the LETG-HRC data of this source, but find that this region is complicated by instrumental feature(s) that we also found in the LETG-HRC observations of 3C~273.

\section{An absorption feature at $\sim 20$~\AA\, -- evidence of the WHIM}

\subsection{Line Identification}

We concentrate on the absorption feature which appears at $20.02~{\rm \AA}$ (619~eV).  This falls at about the same energy where \citet{ckr84} detected a broad ($\sim$ 50-100 eV) absorption feature from an observation with the {\sl Einstein Observatory} Objective Grating Spectrometer. A similar broad feature was detected by \citet{mad91}. However, the line we detect is unresolved and has a width of $\rm <0.039\AA$ (or 1.2 eV) at 90\% confidence. The Poisson significance of that this feature is due to a chance fluctuation is $2.8\times10^{-6}$, equivalent to $4.5\sigma$ for Gaussian statistics.  The probability of finding one such feature by chance in the 1520 bins from 2--40 \AA~is 0.35\%. The 21.61 \AA~feature has a  similar confidence level, while the next most prominent feature has 20\% probability of occuring by chance (these values were confirmed using monte carlo simulations).

Considering cosmic abundances and oscillator strengths (e.g. \citealp{vya95})  for different ions, \ion{O}{8} Ly$\alpha$ is the only strong candidate line between 18 and 20~\AA\,,  the measured wavelength de-redshifted to the source. One scenario which may reconcile \ion{O}{8} Ly$\alpha$  at $18.97{\rm \AA}$ with the $20.02$~\AA\,  of the  detected feature is that this is
the signature of a high velocity $\rm \sim 18,000~km~s^{-1}$ wind/jet blueshifted from the
BL Lac rest frame. In this case the
narrowness of the line would be surprising, especially when compared to
the very broad ($\approx~30,000~\rm km s^{-1}$) feature seen by \citet{ckr84} and \citet{mad91}, although we note that the relativistic jet in the galactic binary SS433 does
have roughly constant velocity \citep{mcs01}.  The implied mass outflow would be hundreds to thousands
of $\rm M_{\odot} \, \rm yr^{-1}$, depending on assumptions about
ionization conditions and beaming factors, as shown in \citet{kkf85}.
They also discuss some of the resulting difficulties with such an outflow.

It is more plausible that the 20~\AA\ absorption is due to \ion{O}{8}
Ly$\alpha$ in a known intervening system at $cz\approx 16,734\rm\
km~s^{-1}$. With the {\sl Hubble Space Telescope} ({\sl HST}), \citet{sps98} discovered a cluster of low metallicity \ion{H}{1}
Ly$\alpha$ clouds along the line-of-sight (LOS) towards PKS~2155-304,
most of which  have redshift between $cz = 16,100\ {\rm km\ s^{-1}}$
and $18,500\ {\rm km\ s^{-1}}$. Using 21 cm images from the {\sl Very
Large Array} ({\sl VLA}), they detected a small group of four
\ion{H}{1} galaxies offset by $\sim 400-800\ {\rm h_{70}^{-1}}$ kpc \footnote{ We use $\rm H_0
= 70h_{70}~km~s^{-1}Mpc^{-1}$, and a standard $\Lambda$CDM model with $\Omega_{m}=0.3$ and $\Omega_{\Lambda} = 0.7$ throughout the paper.}
from the LOS, and suggested that the \ion{H}{1} Ly$\alpha$ clouds
could arise from gas associated with the group.  The velocity centroid
of the four galaxies is $\left<V_{gal}\right> = 16,853\rm~km~s^{-1}$,
which within the uncertainties is identical to that of our absorption
feature.  Therefore the hot gas giving rise to the \ion{O}{8}
absorption can plausibly be associated with the same overdensity
indicated by the \ion{H}{1} Ly$\alpha$ absorbers and the small galaxy
group.

Taking the absorption line to be \ion{O}{8} Ly$\alpha$, the Doppler $b$-parameter must be $b=\sqrt{2}\sigma \rm < 830~km~s^{-1}$ (90\% confidence) since the line is unresolved. If $b > 200\rm~km~s^{-1}$, then the line is unsaturated and $\rm N(O~VIII) \sim 9.5 \times 10^{15}~cm^{-2}$.  At $b < 100\rm~km~s^{-1}$ the line is heavily
saturated, but this can be ruled out by the lack of higher Lyman series
lines in our spectrum. Associating the absorber with the galaxy group
detected by \citet{sps98}, which has a radial velocity dispersion 
of $\left<\sigma_{gal}\right> = 325\rm~km~s^{-1}$, suggests that the unsaturated case is most appropriate, and we adopt that for the subsequent discussion.
However, simulations assuming complete viralization of the hot gas predict $b < 100\rm~km~s^{-1}$ for lines such as this one
\citep{fbc02}, so saturation cannot be ruled out. 

\subsection{Density and Temperature Constraints}

We can constrain the density of the absorbing gas, assuming it is
associated with the intervening galaxy group. First, the upper limit on 
the line width of $\rm \sigma \sim 580~km~s^{-1}$ sets an upper limit to the path length of ${\rm\sim 8h_{70}^{-1}Mpc}$, since otherwise the differential Hubble flow would excessively broaden the line. For $\rm N(O~VIII) \sim 9.5 \times 10^{15}~cm^{-2}$, this gives $n_{b} > (1.0\times10^{-5}\ {\rm cm}^{-3})\
Z_{0.1}^{-1}f_{0.5}^{-1}l_{8}^{-1}$ where $Z_{0.1}$ is the
metallicity in units of 0.1 solar abundance, $f_{0.5}$ is the
ionization fraction in units of 0.5 and $l_{8}$ is the path
length in units of ${\rm 8h_{70}^{-1}Mpc}$. We adopt a metallicity of $\rm 0.1~Z_{\odot}$ \citep{agr89}, which
is the mean value predicted from hydrodynamic simulations for overdensities of $\sim$ 30, with a dispersion of a factor of two \citep{cos99b}. A more reasonable estimate of the 
path length comes from the mean projected separation of $\sim$1 Mpc for the 
galaxies in the group \citep{sps98}, which gives $n_{b} \approx 7.5 \times 10^{-5}\ {\rm cm}^{-3}~Z_{0.1}^{-1}f_{0.5}^{-1}$. In fact, this could be taken as
a plausible upper limit, since the WHIM associated with this overdensity could well be more extended than the galaxies, as are the \ion{H}{1} Ly$\alpha$ clouds.  This implies a range of baryon overdensity over the cosmic mean $\left<n_{b}\right> = 2.14\times 10^{-7}~\rm cm^{-3}$ \citep{tos00} of $\delta_{b} \sim$ 50 - 350. Interestingly, \citet{sps98} estimate an overdensity for the galaxy group of $\delta_{gal}\sim$100.

In the case of pure collisional ionization, temperature is the only parameter 
of importance over a wide range of density so long as the gas is optically
thin.  The \ion{O}{8} ionization fraction peaks at 0.5, and exceeds 0.1
for temperatures $T \sim 2-5 \times10^{6}$ K. Using CLOUDY \citep{fkv98} we find that photoionzation by the cosmic UV/X-ray background is not important for $n_{b} > 10^{-5}\ {\rm cm^{-3}}$.
 
We can compare this estimate for T to what we might expect from the
dynamics of the group of galaxies, if it forms a bound system. Then its 
virial mass is  $\rm M_{vir} \approx 3\times10^{13}h_{70}^{-1}\ M_{\odot}$. Scaling the mass-temperature relationship
down from richer clusters of galaxies \citep{bno98}, gives a temperature of $\sim 9\times 10^{6}$ K. Similarly, using the
relation between velocity dispersion and temperature obtained from fitting a sample of galaxy groups gives a value
of $1-2 \times 10^{7}$ K \citep{xwu00}. These values are somewhat
higher than what we find from the ionization fraction. This could suggest that
the system is not bound and virialized. However,  \citet{mmb96} suggested that the
intragroup gas in spiral-rich groups, such as this one, may be cooler than
that of groups and clusters of early-type galaxies. They suggested that these cooler systems may imprint absorption lines from
 highly-ionized species (e.g. \ion{O}{6}) on quasar spectra, which may
be what we observe.

CLOUDY calculations of the column density ratios between other ions
and \ion{O}{8} give another tight constraint on gas temperature. Based on
the non-detection of other ion species, we calculate the $4\sigma$
upper limits of the column densities of H- and He-like C, N, Ne and He-like
oxygen relative to \ion{O}{8}. N(O~VII)/N(O~VIII) provide the strongest constraints on gas temperature. At roughly $T \ga 10^{6.4}$ K, all the column density ratios satisfy the observed constraints.

Adopting the value of $T = 4-5\times10^{6}$ K, we can compare the
cooling timescale for the  gas to the Hubble time $t_{H}$. The cooling
time scale $t_{cool} \approx (3/2)(kT/n_{b}\Lambda(T))$, here
${\rm \Lambda(T)}$ is the cooling function \citep{sdo93}, which is a function of temperature and metallicity (in
the temperature range of interest, line emission from iron is the most
effective coolant). For 0.1$Z_{\odot}$, the baryon density should be $n_{b} \la
2\times10^{-4}~(t_{H}/t_{cool})~{\rm cm^{-3}}$ so that the gas remains hot. This density limit, however, would be lower by a factor of $\sim
3$ if the abundance is solar. Alternatively, we can use the cooling
argument to set an upper limit on the metallicity. Adopting a
baryon density of ${\rm 7.5\times10^{-5}\ cm^{-3}}$ and requiring $t_{cool} > t_{H}$, the cooling function $\Lambda$ must be less than ${\rm 2\times 10^{-23}\ ergs\ cm^{3}s^{-1}}$, which implies a metallicity of $\la 0.3Z_{\odot}$.

Following \citet{lmw91} and \citet{tsa00} we
estimate the baryonic content $\rm \Omega_{b}(O~VIII)$ that can be probed by our
\ion{O}{8} resonance absorption line, expressed in units of $\left<n_{b}\right>$. We assume  conservative upper limits of $\rm Z
\la 0.5Z_{\odot}$ and \ion{O}{8} ionization fraction of $f \la
0.5$. Given the path length of $\Delta z \la 0.116$, we estimate
$\rm \Omega_{b}(O~VIII) \ga 0.005h_{70}^{-1}$. This is about 10\% of the
total baryon fraction, or about 30-40\% of the WHIM gas, if the WHIM gas contains about 30-40\% of total baryonic matter \citep{dco01}. This baryon fraction is consistent with the prediction from \citet{plo98} based on a simple analytic model; however, these numbers are very rough estimates, due to the large uncertainties in
the abundance and ionization fraction.

\section{Discussion \& Summary}

Our detection of \ion{O}{8} Ly$\alpha$ from an intervening overdense region in the
IGM at $cz \sim 16,600\rm~km~s^{-1}$, together with the detection by  
\citet{nic02} of 
absorption from systems at $cz \sim 0$ associated with the Galaxy or local group, have begun
to reveal the much-anticipated warm-hot component of intergalactic matter
(\citealp{hgm98,cos99a,dco01,fbc02}).

The \ion{O}{8} Ly$\alpha$  feature we observe is plausibly associated with the same
cosmic overdensity that gives rise to the \ion{H}{1} Ly$\alpha$ clouds
and the four spiral galaxies found by \citet{sps98}. Three of the
galaxies have radial velocities within $\sim100~\rm km~s^{-1}$ of one another,
while the fourth is within $\sim400~\rm km~s^{-1}$.  In contrast, the velocities
of the \ion{H}1 Ly$\alpha$ clouds span $\sim1800~\rm km~s^{-1}$, with two
falling very close to the velocities of the closest projected galaxies. The
galaxies represent an overdensity of $\delta_{gal} \sim 100$ relative to the mean. Our
estimates for the overdensity of the hot absorbing gas are $\delta_{b} \sim$50-350. The
\ion{H}{1} Ly$\alpha$ clouds have very low metallicity, indicating that
they are primarily composed of primordial material, whereas the hot
gas should have an oxygen abundance of $\sim 0.1$ solar to give sensible
values for the column density. It is unlikely
that the same gas is responsible for both \ion{H}{1} and \ion{O}{8} absorption. Following \citet{sps98}, we
identify this as a region of modest overdensity (the galaxy group) connected
to filamentary or sheet-like structures demarcated by the \ion{H}{1} Ly$\alpha$ clouds, such as those seen in numerous
cosmological simulations.

We note that the detection of a system such as this is roughly consistent with
previous simulations.  Based on the models of \citet{plo98} and \citet{fbc02}, we estimate the cumulative absorption line number along a random sightline (Fig.~2). We adopt the numerical relation between baryon overdensity and metallicity from \citet{cos99b} and assume collisional ionization. This observation is $\sim 1.5\sigma$ above the predicted value at the measured column density. According to Fig.~3, the probability of detecting such an \ion{O}{8} Ly$\alpha$ line or stronger to $z\sim 0.1$ is $\sim$ 10\%. The consistency, if borne out by further studies, suggests that the simulations
are valid descriptors of the warm/hot component of the IGM, at least
for overdensities of $\sim$100.
Using those simulations, we estimate that we can probe about 10\% of 
total baryons and about 30-40\% of the WHIM gas in the local universe. 

Other observations of bright AGN 
with the {\sl Chandra} and {\sl XMM-Newton} grating spectrometers are likely
to reveal additional WHIM absorption features, and may turn the present
few samplings into a real forest of high ionization Ly$\alpha$ or He$\alpha$ lines. However, the
sensitivity of these instruments permit us only to probe the high density
tail of the distribution (e.g. \citealp{fca00}). It will be left to future missions like {\sl Constellation-X} and  {\sl XEUS}, assuming they have both high throughput and 
spectral resolving powers of 1000 or more at energies 0.1-1 keV, to fully reveal the X-ray forest.

We are grateful to F.~Nicastro and his colleagues for sharing an advance
copy of their paper.  We thank the other members of the MIT/CXC team for its support. We also thank the referee, J.~M.~Shull, for helpful comments. This work is supported in part by contracts NAS 8-38249 and SAO SV1-61010.

\clearpage

\vbox{
\scriptsize
\begin{center}

\begin{tabular}{lllll}
\multicolumn{5}{c}{Table 1: Fitting parameters of the X-ray absorption Lines~~~} \\
\hline
\hline
& O~{\sc viii} Ly$\alpha$ & O~{\sc viii} Ly$\alpha$ &  O~{\sc vii} He$\alpha$ &  O~{\sc vii} He$\beta$ \\
\hline
$\lambda_{rest}$$^{a}$      & 18.9689  & 18.9689 & 21.6019 & 18.6288 \\
$\lambda_{obs}$       & $18.95_{-0.02}^{+0.02}$       & $20.02_{-0.015}^{+0.015}$            & $21.61_{-0.01}^{+0.01}$       & $18.63_{-0.02}^{+0.02}$ \\ $cz({\rm km\ s^{-1}})$ & $-300_{-315}^{+317}$  & $16624\pm237$      & $112_{-138}^{+140}$  & $20_{-323}^{+321}$ \\ 
Line Width$^{b}$      &    $ ... $           &  $< 0.039$                & $<0.027$             & $ ... $\\ 
$\rm Line\ Flux^{c}$   & $3.0_{-2.3}^{+1.2}$  & $4.8_{-1.9}^{+2.5}$       & $5.5_{-1.7}^{+3.0}$  & $1.1_{-0.6}^{+1.8}$ \\ 
EW (m${\rm \AA}$)      & $8.8_{-6.9}^{+3.5}$  & $14.0_{-5.6}^{+7.3}$      & $15.6_{-4.9}^{+8.6}$ & $3.3_{-1.8}^{+5.5}$ \\ 
SNR$^{d}$              & 3.0                  & 4.5                       & 4.6                  & 1.1 \\
\hline
\end{tabular}

\parbox{3.5in}{
\vspace{0.1in}
\small\baselineskip 9pt
\footnotesize
\indent
a. Rest-frame wavelengths in units of $\rm \AA$ \citep{vya95}.\\
b. 90\% upper limit of the line width $\sigma$, in units of $\rm\AA$.\\
c. Absorbed line flux in units of $\rm 10^{-5}~photons~cm^{-2}s^{-1}$.\\
d. In units of equivalent sigma of a Gaussian distribution with the same confidence level as the Poisson significance gives.\\
}
\end{center}
\normalsize
\centerline{}
}

\clearpage
\figcaption[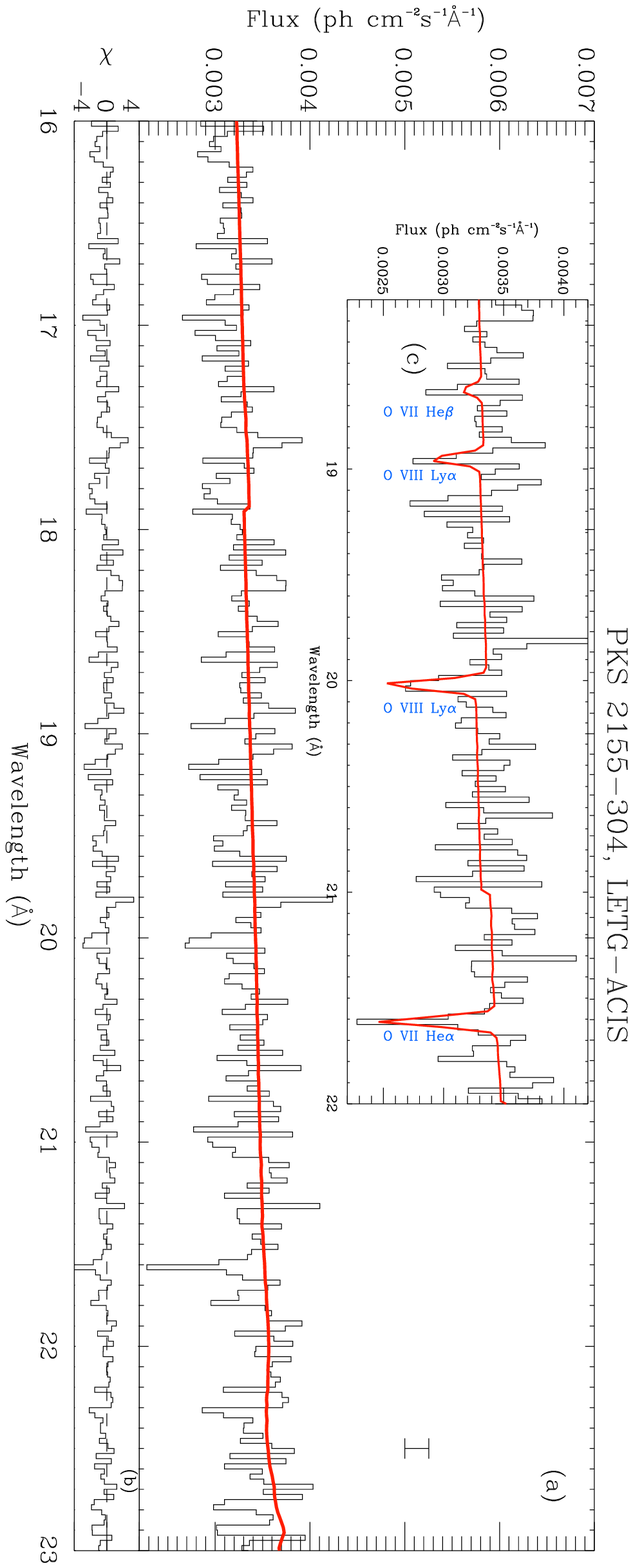]{The {\sl Chandra} LETG-ACIS spectrum of PKS~2155-304 between 16 and 23 $\rm\AA$. (a): The red solid line is the fitted continuum (Galactic absorbed power law plus the polynomial), the averaged $1\sigma$ error bar plotted on the right is based on statistics only; (b) the signal-to-noise ratio; (c) enlarged portion of the spectrum between 18 and 22 $\rm\AA$, the red solid line is the fitted continuum plus three Gaussian models, the ion species are labeled in blue.\label{fig1}}

\figcaption[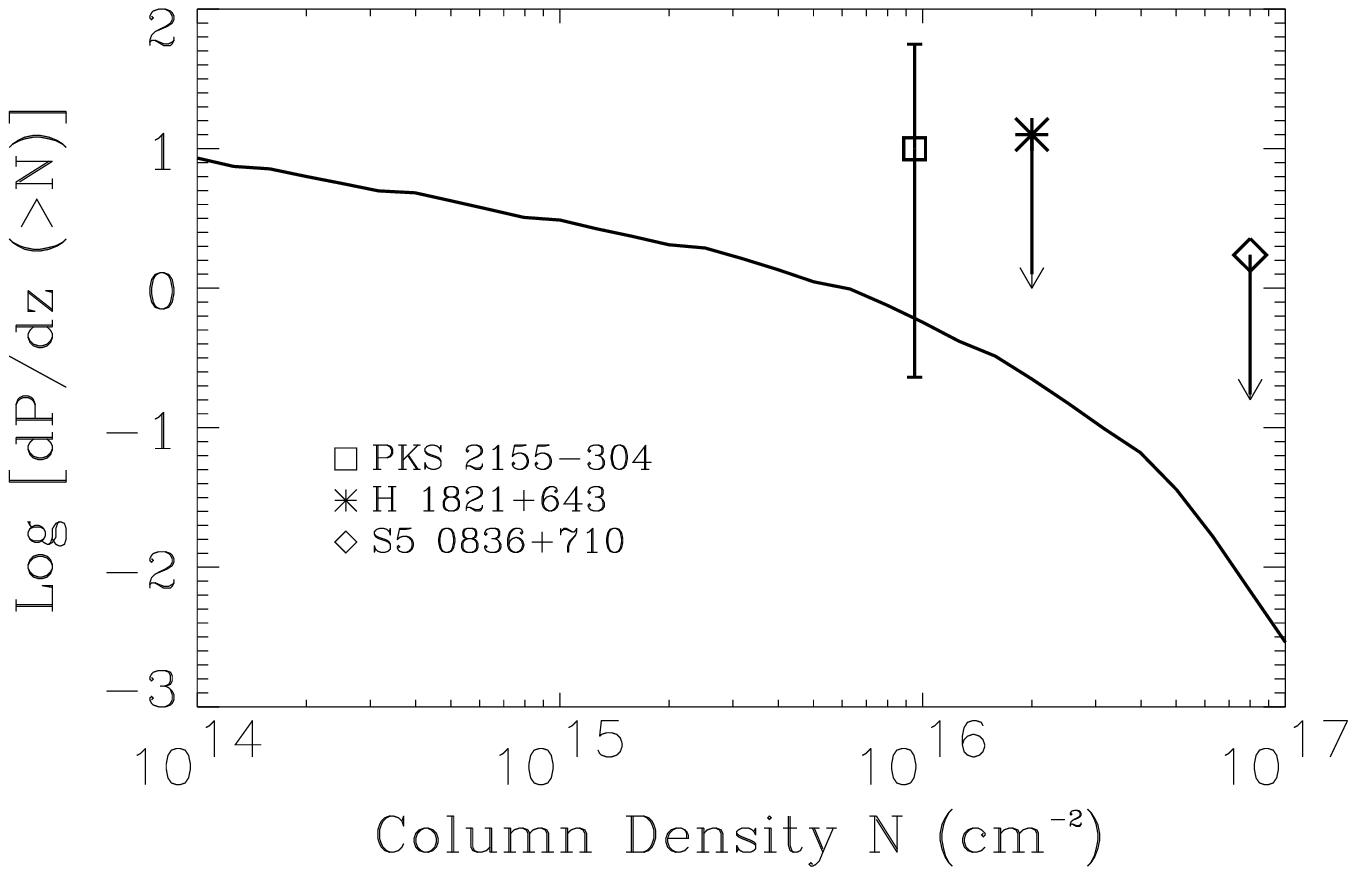]{Cumulative \ion{O}{8} absorption lines per unit redshift vs. column density. Three data points are: PKS~2155 (this paper), H~1821 \citep{fdl02}, S5~0836 \citep{fmb01}. The errors are estimated at $2\sigma$ level \citep{geh86}. For non-detection cases, $2\sigma$ upper limits are given.}

\clearpage
\begin{figure}
\epsscale{0.5}
\plotone{f1.eps}
\end{figure}
\clearpage

\begin{figure}
\epsscale{1.0}
\plotone{f2.eps}
\end{figure}

\end{document}